\documentclass[twocolumn,showpacs,amsmath,amssymb]{revtex4}

\usepackage{graphicx}
\usepackage{dcolumn}
\usepackage{bm}
\newcommand{\etal}{\emph{et al.}}
\newcommand{\be}{\begin{equation}}
\newcommand{\ee}{\end{equation}}
\newcommand{\bfig}{\begin{figure}}
\newcommand{\efig}{\end{figure}}
\newcommand{\incl}{\includegraphics}

\begin{document}

\title{The Lorenz number in CeCoIn$_5$ inferred from the thermal and charge Hall currents 
}
\author{Y. Onose$^1$, N. P. Ong$^1$ and C. Petrovic$^2$
}
\affiliation{
$^1$Department of Physics, Princeton University, Princeton, NJ 08544, USA\\
$^2$Department of Physics, Brookhaven National Laboratory, Upton, N.Y. 11973, USA
}

\date{\today}      

\begin{abstract}
The thermal Hall conductivity $\kappa_{xy}$ and 
Hall conductivity $\sigma_{xy}$ in CeCoIn$_5$ are used to determine the Lorenz number ${\cal L}_H$ at low temperature $T$.  This enables the separation of the observed thermal conductivity into its electronic and non-electronic parts.  We uncover evidence for a charge-neutral, field-dependent thermal conductivity, which we identify 
with spin excitations.  At low $T$, these excitations dominate the scattering of charge carriers.  We show that suppression of the spin excitations in high fields leads to a steep enhancement of the electron mean-free-path, which leads to an interesting scaling
relation between the magnetoresistance, thermal conductivity and $\sigma_{xy}$.
\end{abstract}

\maketitle

\section{Introduction}
The heavy-electron system CeCoIn$_5$ exhibits a host of unusual electronic properties 
of current interest.  In the superconducting state, strong
evidence for $d$-wave pairing symmetry has been reported~\cite{Petrovic,Movshovich,Izawa,Ormeno,Aoki}. 
The FFLO state involving pairing with unequal spin populations 
in an in-plane magnetic field ($\bf H\perp c$) has been proposed ~\cite{FFLO}.  
The phase diagram in a perpendicular field ($\bf H || c$) has also received
wide attention~\cite{Paglione,Bianchi}.  In an extended region in $T$-$H$ plane 
surrounding the superconducting (SC) state (labelled I in Fig. \ref{kT}a), 
the resistivity $\rho$ and heat capacity exhibit distinctive ``non-Fermi liquid'' characteristics:
$\rho\sim T$~\cite{Paglione}, while the Sommerfeld parameter $\gamma(T)\sim \log T$ ~\cite{Bianchi}.  
When $H$ exceeds the boundary $H_s(T)$, $\rho$ recovers the Fermi-liquid
form $\rho = \rho_0 + AT^2$ and the unconventional features of $\gamma(T)$ 
are suppressed.  The high-field region (labelled II in Fig. \ref{kT}a)
is called the Fermi-liquid region.  The boundary field $H_s(T)$ and the
upper critical field $H_{c2}$ terminate at a quantum critical point (QCP) 
as $T\rightarrow 0$ (the field scale $H_Z$ is discussed below). 
Several parameters characterizing resistivity display divergent behavior as the QCP is approached~\cite{Paglione}.
In addition, a large Nernst signal is observed in I~\cite{Bel,Onose}.  

\bfig  
\incl[width=8.8cm]{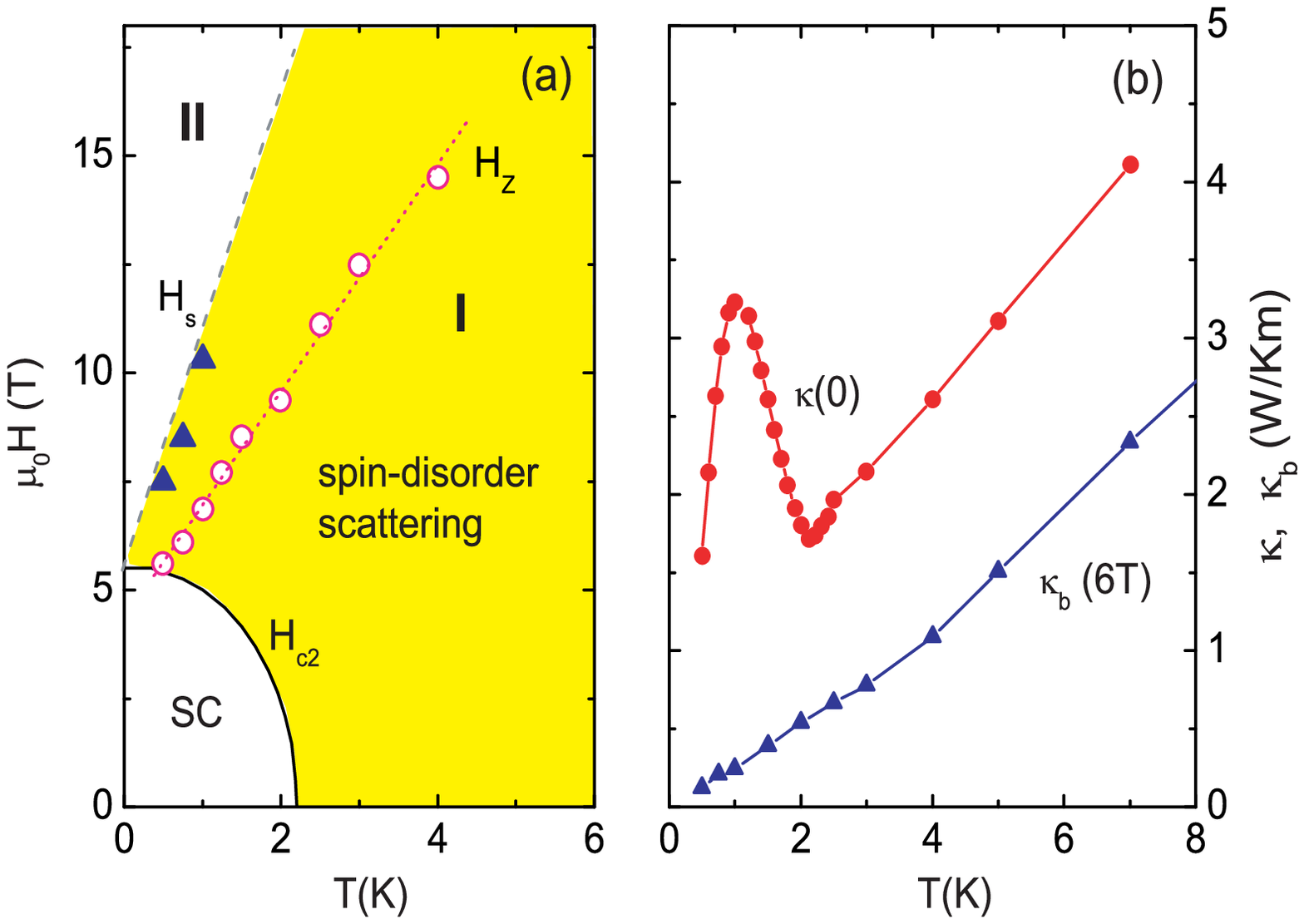}
\caption{\label{kT}
(a)  The phase diagram of CeCoIn$_5$ adapted from Refs. \cite{Paglione,Bianchi},
showing some features inferred from our experiment.
The dashed line is the boundary $H_s(T)$ between the Fermi liquid (II) and non-Fermi 
Liquid (I) regions reported in Refs \cite{Paglione,Bianchi}. 
Solid triangles are our values for $H_s(T)$.  
Open circles indicate the field scale $H_Z$ (this work) 
above which the current ratio ${\cal R}_{\sigma} = Z$ (see Eq. \ref{Rs}).
(b) The $T$ dependence of $\kappa$ in zero field (solid circles)
showing a large qp peak below $T_c$.  The background term $\kappa_b$ 
measured at 6 T is shown as solid triangles.  Below $\sim$8 K, $\kappa_b$ is 
largely comprised of a term $\kappa_s$ that is very 
field dependent (and identified with spin excitations).
}
\efig

To clarify the electronic state in the region I, we have measured extensively 
the in-plane thermal conductivity $\kappa\equiv \kappa_{xx}$ with $\bf H||c||z$,
and the thermal Hall conductivity $\kappa_{xy}$ (the Righi-Leduc effect)
in crystals with very long electron mean-free-path $\ell$.  
In addition, we measured the electrical
conductivity $\sigma\equiv\sigma_{xx}$ and Hall conductivity $\sigma_{xy}$.
The crystals, grown from metallic flux~\cite{Petrovic}, are plate-like with the $c$ axis
normal to the broadest faces and the $a$ axis along one edge (for 
structure, see Ref. \cite{Kalychak}).  
The zero-field $\kappa$ displays a prominent peak
below $T_c$ (solid circles in Fig. \ref{kT}b), which arises from the steep increase
in $\ell$ of Bogolyubov excitations in the superconducting state.  
Using the Hall conductivities $\kappa_{xy}$ and $\sigma_{xy}$, 
we demonstrate the validity of the Wiedemann-Franz (WF) law, and then 
use the Lorenz number to separate the total $\kappa$ into 
its electronic and non-electronic components $\kappa_e$ and $\kappa_b$,
respectively.  From the strong field dependence observed in $\kappa_b$,
we infer that spin excitations provide the dominant scattering
mechanism for the charge carriers in the region I.  The application of
an intense field leads to suppression of this scattering channel
and a sharp increase in $\ell$.  This insight sheds light on the
large magnetoresistance and the unusual features of the Hall effect.
We discuss the implications for Cooper pairing in the SC region.

\section{Thermal and charge conductivity tensors}
The thermal resistivity tensor $W_{ij}$ is measured by applying a weak gradient 
($\delta T\sim$ 10 mK along length of the crystal at 0.5 K).  Below 2 K, the
strong variation of $\ell$ with $T$ and $H$ is potentially the largest source of error 
in comparing $W_{ij}$ (measured in finite $\delta T$) with $\rho_{ij}$ ($\delta T$ = 0).
We minimized the uncertainties by extensive calibration of the 
RuO$_x$ thermometers (glued to the crystal), and using very slow field scans 
(0.1-0.2 T/min. at 0.5 K).  

We emphasize that, because $\ell$ attains very large values 
below 10 K, it is necessary to use the full matrix inversion to 
reliably convert the measured tensors $W_{ij}$ and $\rho_{ij}$ into 
their reciprocal conductivity tensors, 
e.g. $\kappa_{xx} = W_{xx}/(W_{xx}^2+W_{xy}^2)$.  Experimentally, this means that 
$\kappa$ and $\sigma$ in strong fields cannot be obtained 
without measuring simultaneously the diagonal and off-diagonal elements of $W_{ij}$ 
and $\rho_{ij}$ (leaving out the Hall tensor elements 
leads to errors in $\kappa$ and $\sigma$ of 30$\%$ or more at low $T$).

\bfig [h]  
\incl[width=7.5cm]{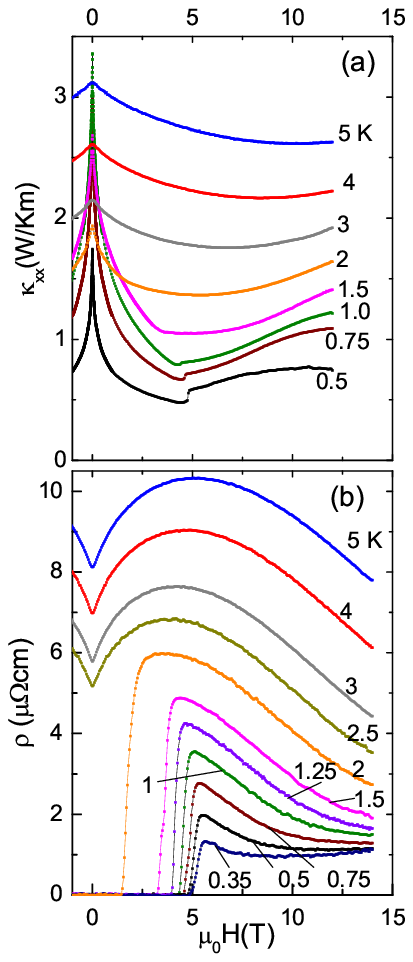}
\caption{\label{krho}
(a) Curves of $\kappa_{xx}$ vs. $\bf H||c$ at selected $T$.  
The steep suppression of the zero-field anomaly produces a sharp spike at each $T<T_c$. 
(b) The magnetoresistance (MR) $\rho$ vs. $\bf H||c$ showing the strong negative MR above
4 or 5 T. 
}
\efig

Figure \ref{krho} compares the field dependences of $\kappa$ (Panel a) and the in-plane 
resistivity $\rho$ (Panel b) at temperatures from 5 K to 0.5 K 
with $\bf H||c$.  Above 5 K, an increasing field decreases slightly the
observed $\kappa$.  With decreasing $T$, however, this trend changes.
At 2 K, $\kappa$ rises gradually with $H$.  At even lower $T$ (0.5--1.5 K), 
this rising trend becomes firmly established in the normal state 
when $H$ exceeds $H_{c2}$ (step in $\kappa$).  
In the SC region below $H_{c2}$, a prominent feature in $\kappa$ is the sharp, narrow spike which represents the rapid field suppression of the zero-field peak caused by
scattering of Bogolyubov excitations from vortices~\cite{Krishana,Zhang01,Durst}.
The spike in $\kappa$ vs. $H$ below $T_c$ is much larger than 
previously reported~\cite{Izawa}.  This reflects a much longer $\ell$ in 
the present samples.

The complicated behavior of $\kappa(T,H)$ arises 
because it is the sum of the electronic term $\kappa_e$ 
and a ``background" term $\kappa_b$ carried by 
charge-neutral excitations (spin excitations
and phonons), viz.  
\be
\kappa(T,H) = \kappa_e(T,H) + \kappa_b(T,H).
\label{k}
\ee

Our main finding is that, in CeCoIn$_5$, the charge-neutral term
$\kappa_b(T,H)$ displays an unexpectedly strong $H$ dependence.  
Its $T$-profile at $H$ = 6 T is shown as solid triangles in Fig. \ref{kT}b.

As previously found~\cite{Paglione,Nakajima}, CeCoIn$_5$ exhibits 
a large magnetoresistance (MR) (Fig. \ref{krho}b).  
The initial positive MR ($H<$3 T) is caused by suppression
of superconducting fluctuations which we discuss elsewhere~\cite{Onose}.  Our focus is on the 
negative MR that prevails for $H>$4 T at all $T$ below $\sim$30 K.  
As $T$ decreases towards $T_c$, the negative MR becomes 
pronounced.  At 2 K, $\rho$ decreases by $\sim$2.5 when $H$ reaches 14 T.  
Both the sign and magnitude preclude classical MR associated 
with the Lorentz force.  As shown below, the MR results from 
a steep enhancement of $\ell$ with increasing field.

\section{Lorenz number from scaling of $\kappa_{xy}/T$ to $\sigma_{xy}$}
The field enhancement of $\ell$ strongly influences the field profiles of the heat and charge
currents.  To disentangle these effects, we exploit the WF law, which states that 
the ratio $\kappa_e/T\sigma$ is close to the Lorenz factor 
${\cal L}_0 = \frac13\pi^2(k_B/e)^2$.  In the elemental metals, the WF law 
is nearly universally obeyed at 300 K as well as in the impurity-scattering 
regime below 4 K, while deviations are common in between.  
However, in many interesting metals with low carrier 
densities, $\kappa_e$ cannot be measured directly because the 
charge-neutral term $\kappa_b$ (usually from phonons) is comparable in size or larger.  

Recently, a way to separate $\kappa_e$ from $\kappa$ using the 
Righi-Leduc effect was introduced. 
Zhang \etal~\cite{Zhang00} have shown that the Lorenz ratio may 
be determined from the ratio ${\cal L}_H \equiv \kappa_{xy}/T\sigma_{xy}$.
(Essentially, the Righi-Leduc effect senses only the electronic entropy
current, while filtering out the charge-neutral components which
do not have a Hall response.  Since the latter also do not contribute
to $\sigma_{xy}$, the ratio of the 2 Hall currents yields the 
WF ratio. The WF-Hall method was tested on high-purity Cu and 
applied to cuprates~\cite{Zhang00}.)  CeCoIn$_5$ is well-suited for this 
method because the 2 Hall conductivities are large.

\bfig[h] 
\incl[width=8cm]{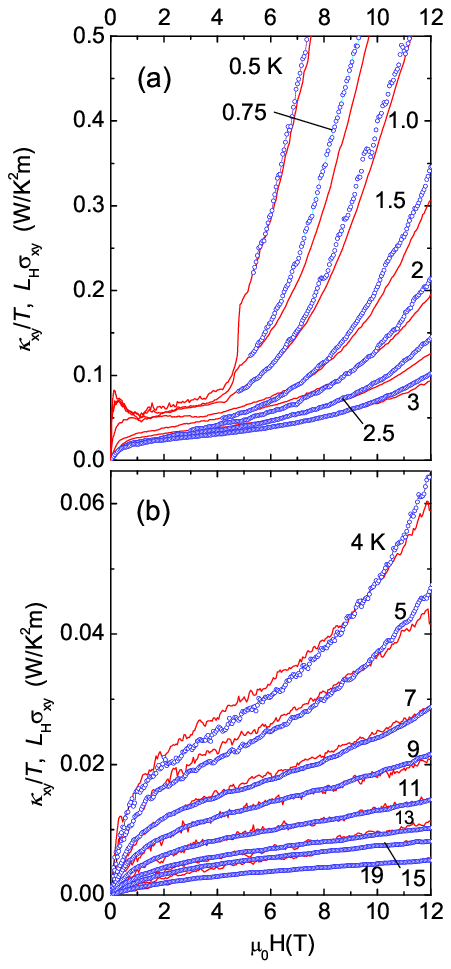}
\caption{\label{WF}
Comparison of the thermal Hall conductivity $\kappa_{xy}/T$ (solid curve) 
and the Hall conductivity $\sigma_{xy}$ (open circles) scaled by 
an $H$-independent ${\cal L}_H$ at $T\le$ 3 K (Panel a) and $T>$ 3 K (Panel b).  
Below 1.5 K (Panel a), $\kappa_{xy}/T$ displays a peak anomaly in weak $H$.
Both Hall currents are electron-like in sign.  
}
\efig

\bfig[h] 
\incl[width=7cm]{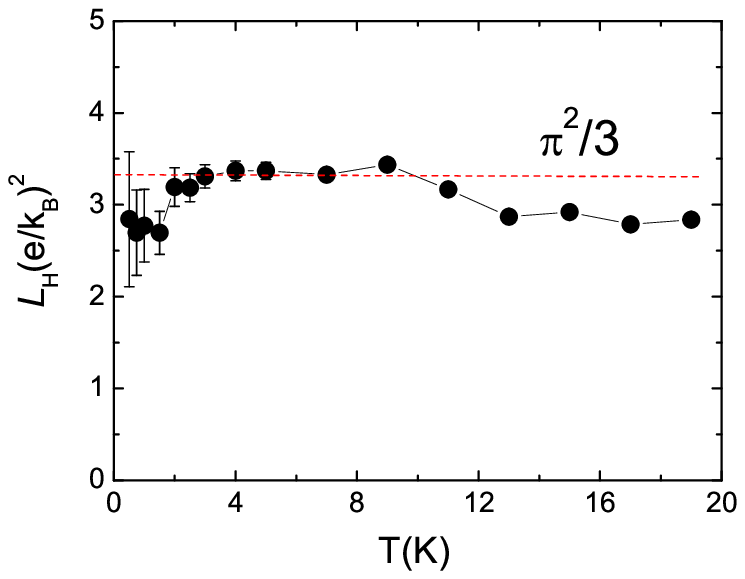}
\caption{\label{Lorenz}
The Lorenz number ${\cal L}_H$ 
obtained by scaling $\kappa_{xy}/T$ to $\sigma_{xy}$
in Fig. \ref{WF}.  ${\cal L}_H$ is plotted 
in units of $(k_B/e)^2$.  The dashed line is the 
Sommerfeld value $\pi^2/3$.
}
\efig

The Hall resistivity in CeCoIn$_5$ was previously reported~\cite{Nakajima},
but it is the Hall conductivity that is of interest here.  At each $T$, we find that the 
profile of $\sigma_{xy}$ vs. $H$ matches that of $\kappa_{xy}$ even when the 
two curves display strong curvature vs. $H$.  
The curves of $\kappa_{xy}/T$ and ${\cal L}_H\sigma_{xy}$ 
are plotted together in Fig. \ref{WF} for $T\le$ 3 K (Panel a), and $T>$ 3 K (b).  
Let us first note that the curves share 2 characteristics rarely seen in 
Hall experiments.  In weak $H$, the curves rise from zero with strong negative curvature 
to produce a knee-like feature.  In addition, the curvature changes its sign to positive in higher fields; both Hall conductivities increase more rapidly than the first power in $H$ in strong fields.  Further, we note the peak anomaly in weak $H$ shown by $\kappa_{xy}$ 
below 1.5 K.  We return to these unusual features later.

At each $T$, $\kappa_{xy}/T$ and ${\cal L}_H\sigma_{xy}$ may be scaled 
together over the entire field range by adjusting ${\cal L}_H$. 
We emphasize that ${\cal L}_H$ is an $H$-\emph{independent}
scaling parameter (otherwise, it does not make sense to discuss
scaling between $\kappa_{xy}/T$ and $\sigma_{xy}$).
In view of the pronounced nonlinearity, the close match between 
the 2 field profiles is strong evidence that the WF law is valid
with a field-independent Lorenz number.  
The inferred values of ${\cal L}_H$ are 
plotted in Fig. \ref{Lorenz}.  Between 2 and 10 K,
${\cal L}_H/(k_B/e)^2$ is close to the Sommerfeld value $\pi^2/3$,
but seems to deviate slightly downwards below 2 K.

\section{Separation of electronic and non-electronic heat currents}
We next determine $\kappa_e(T,H)$ and $\kappa_b(T,H)$ in
Eq. \ref{k}.  Using the values of ${\cal L}_H$ in 
Fig. \ref{Lorenz}, we convert the measured $\sigma$ into 
$\kappa_e(T,H)$ via
\be
\kappa_e(T,H) = T\sigma(T,H){\cal L}_H(T).
\label{ke}
\ee
Subtracting the curve of $\kappa_e(T,H)$ from $\kappa(T,H)$ at each $T$, we finally
determine $\kappa_b(T,H)$, which is plotted in Fig. \ref{kb}a.  

At low $T$, $\kappa_b$ is found to be strongly $H$ dependent.  
In general, $\kappa_b$ is the sum of the spin-excitation conductivity $\kappa_s$ and the phonon conductivity
$\kappa_{ph}$, viz. $\kappa_b(T,H) = \kappa_s(T,H)+\kappa_{ph}(T)$.  
When spin-disorder scattering of phonons is important, an applied field
generally leads to an \emph{increase} in $\kappa_{ph}$ because $H$ 
suppresses spin disorder, which is opposite to what is observed.  
Consequently, we identify all the
field dependence with the spin-excitation term $\kappa_s(T,H)$.
Figure \ref{kb}a shows that $\kappa_s$ accounts for a 
large fraction of $\kappa_b$ between 5 K and 1 K.  At the lowest $T$ (0.5--1.5 K), the curve
of $\kappa_b$ falls to a floor value at the
field scale $H_s(T)$, which is observable as a break-in-slope.  
In the phase diagram in Fig. \ref{kT}a, $H_s(T)$ is 
seen to lie close to the I/II boundary (solid triangles). With the present 
data, we cannot determine $H_s$ above 1.5 K.  Nonetheless, 
the trends of the curves in Fig. \ref{kT}a suggest that,
throughout the region I up to 5 K, $\kappa_b$ is greatly 
reduced from its zero-field values when $H$ reaches
the I/II boundary in the phase diagram (Fig. \ref{kT}a).

We compare our results with Ref.~\cite{Paglione2}, where the Lorenz number was found
by estimating the phonon term $\kappa_{ph}$ from measurements on a La-doped
sample~\cite{CeRhIn}.  
The authors compute $\kappa_e$ 
as the difference $\kappa(H,T) - \kappa_{ph}(T)$, assuming $\kappa_{ph}$ 
to be strictly $H$-independent.  The inferred Lorenz number ${\cal L}_{eff}$ 
is reported to be suppressed by $\sim 30\%$ in a 10-T field~\cite{Paglione2}, 
in contrast with our $H$-independent Lorenz number.  
The field dependence of ${\cal L}_{eff}$ in Ref.~\cite{Paglione2} 
comes from identifying $\kappa_b$ entirely with $\kappa_{ph}$ 
which is assumed $H$-independent. However, we note that the strong $H$ dependence of $L_{eff}$~\cite{Paglione2} is mainly observed below our temperature range.  
Differences at higher $T$ may stem from identifying bosonic contribution
through experiments on doped samples~\cite{Tanatar}, as opposed to the Hall response
here.  In both cases, the Lorenz ratio attains the Sommerfeld value above 4 K.

\bfig[h]  
\incl[width=8cm]{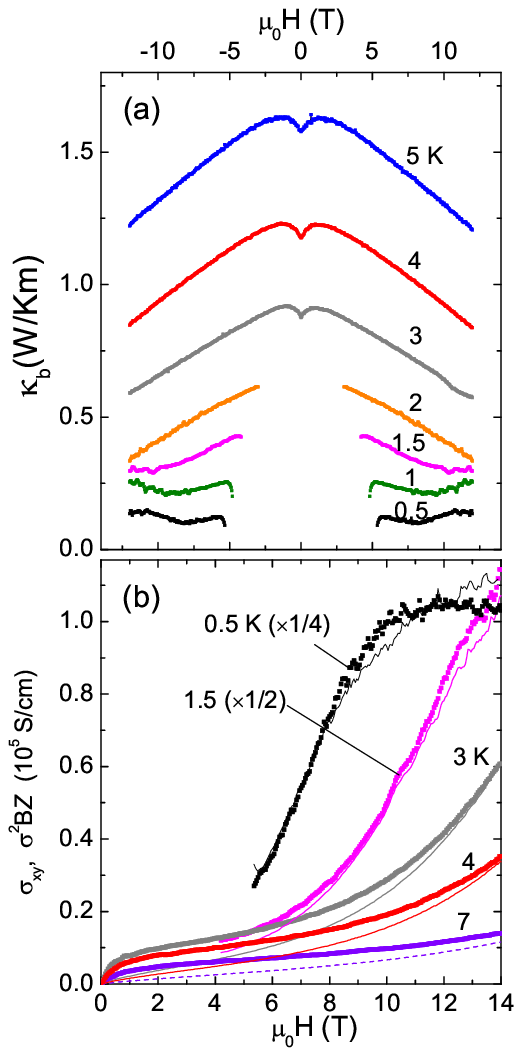}
\caption{\label{kb}
(a) The curves of $\kappa_b = \kappa_s + \kappa_{ph}$ vs. $H$ 
obtained by subtracting $\kappa_e$ 
from the observed $\kappa$ at each $T$ ($\kappa_b$ is not obtained
below $H_{c2}$ because $\rho$ = 0).
At 0.5, 0.75 and 1 K, the $H$ dependence
shows a kink at $H = H_s(T)$ at the boundary between I and II (Fig. \ref{kT}a). 
The curve of $\kappa_b$ vs. $T$ at 6 T is shown in Fig. \ref{kT}b.
(b) Comparison of $\sigma_{xy}$ (bold curves) with the quantity $\sigma^2B{\cal Z}$ 
(thin curves).  Above the field $H_Z(T)$, the 2 quantities match over 
nearly 2 decades with one scaling constant ${\cal Z} = 1\times 10^{-7}$ cm$^3$/C.  
Below $H_Z$, however, the scaling is spoilt by an ``excess Hall current".
}
\efig

\section{Spin degrees and charge transport}
In a conventional magnet, an external $\bf H$ raises the magnon dispersion energy which
reduces the spin-wave population and their thermal conductivity.
In the region I of CeCoIn$_5$, the uniform susceptibility $\chi$ is strongly 
enhanced, but conventional long-range magnetic order 
seems to be absent.  However, in heavy fermions, spin-ordered states 
involving the local moments in the $f$ bands are widely postulated.  
Incipient spin ordering may exist above $T_c$ in CeCoIn$_5$
(see Broholm~\cite{Broholm}).
Although our analysis is guided by the known properties 
of conventional spin waves, a more exotic kind of spin ordering is not precluded, and $\kappa_s$ may 
derive from spin-excitations in unconventional spin-ordered states.  
Because of hybridization between the $f$ and $s$-$p$-$d$ states and large spin-orbit
coupling, the spin excitations will strongly scatter the charge carriers.  
We write $\kappa_s  = \frac13 c_sv\lambda$ where $c_s$ is the heat capacity 
of the spin excitations, $v$ the average velocity and $\lambda$ their mean free path.
As the local moments align with $\bf H$, the spin excitation population $n_s\sim c_s$
decreases steeply in field.  We interpret the curves of $\kappa_b$ in Fig. \ref{kb}a 
as the sharp field-suppression of $n_s$ at low $T$.  While evidence for heat currents 
carried by spin excitations have been reported for low-dimensional 
oxides~\cite{1D}, a distinguishing feature in CeCoIn$_5$ is that
changes in $\kappa_s$ strongly affect the charge currents, which we describe next.

The curves in Fig. \ref{kb}a reveal that the charge-neutral 
conductivity $\kappa_b$ decreases strongly with increasing $H$.
This trend is opposite to that in the electronic conductivity
$\sigma$.  As the former decreases, the latter rises in 
almost direct proportion.  
[The ratio of $\kappa_b(H)$ evaluated at $H$ = 0 and 14 T, 
$\kappa_b(0)/\kappa_b(14)\sim$1.5 and 2.7 at 
5 K and 2 K, respectively.  These ratios match the
corresponding ratios $\rho(0)/\rho(14)$ at the
same $T$ in Fig. \ref{krho}b.]
Panel b of Fig. \ref{kb} shows the 
steep increase of $\sigma^2$ 
and $\sigma_{xy}$ with $H$.  As discussed above,
the sharp decrease in $\kappa_b$ with $H$ reflects a decrease in the density 
of spin excitations $n_s$.  Hence the correlated increase in
$\sigma$ implies that spin excitations are the dominant scattering mechanism of the carriers at these temperatures.  The steep increase in $\ell$ in high field
is caused by field-suppression of the spin excitations.
Full suppression of this scattering channel, attained when $H$ reaches $H_s$, leads 
to the unusually large $\ell$ in the region II.

\section{Magnetoresistance and current ratio ${\cal R}_{\sigma}$}
We have also found that the strong enhancement of $\ell$ 
forges a link between the unusual MR and Hall effect.  The observed $\sigma_{xy}$ is the sum of contributions $\sigma_{xy}^i$ from each FS sheet $i$.  
Assuming that the lifetime $\tau_i$ on each sheet is dominated by spin-disorder
scattering, all $\tau_i$ follow the same monotonically rising function 
of field $g(B)$.  As a result, we have $\sigma_{xy}\sim Bg(B)^2$, 
while the conductivity $\sigma\sim g(B)$: The Hall current grows in 
direct proportion to the square of the longitudinal current 
multiplied by $B$.

\bfig[t]  
\incl[width=6cm]{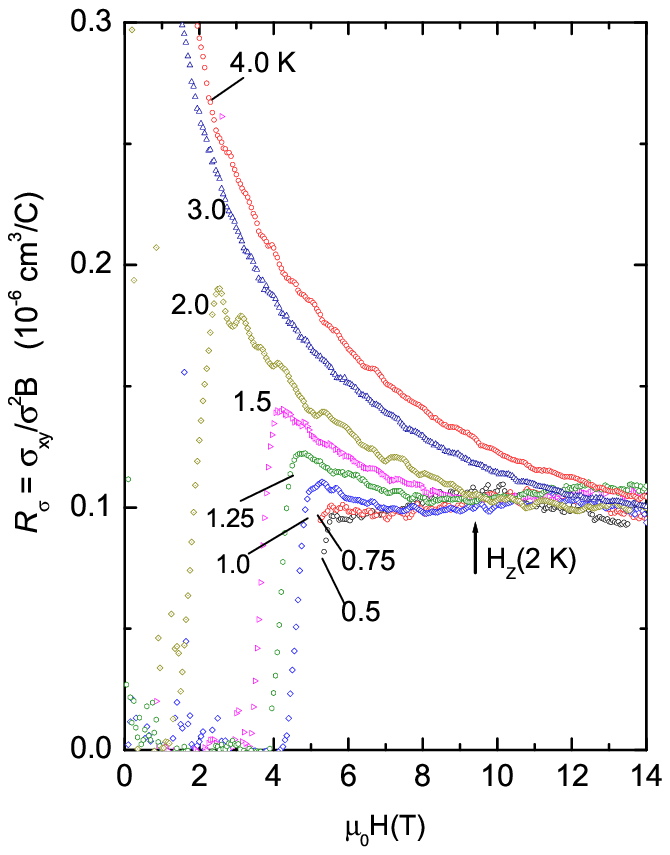}
\caption{\label{R}  Curves of the current ratio ${\cal R}_{\sigma}$ (Eq. \ref{Rs})
at selected $T$.  Above $H_Z(T)$, ${\cal R}_{\sigma}$ attains a plateau value ${\cal Z}$ which is
$T$-independent at our resolution (arrow indicates $H_Z$ at 2 K).  
The steep variation in weak $H$ reflects the excess
Hall current discussed in Fig. \ref{kb}b.
}
\efig

To test this assumption, we compare the curves of $\sigma_{xy}$ 
with $\sigma^2B$ at low $T$ (Fig. \ref{kb}b).
In high fields, the quantity $\sigma^2B{\cal Z}$ 
(thin curves) can be made to match $\sigma_{xy}$ (bold curves) by setting the 
$T$-independent constant ${\cal Z} = 1\times 10^{-7}$ cm$^3$/C.  
The match is excellent for fields above
a cross-over field $H_Z(T)$.  As $H$ decreases below $H_Z$, 
however, $\sigma_{xy}$ increasingly exceeds $\sigma^2B{\cal Z}$.  
The negative curvature (knee) feature described earlier now 
appears as a small ``excess" Hall current below $H_Z$.  

This high-field scaling is made more apparent if we plot the quantity
\be
{\cal R}_{\sigma}(T,B) = \sigma_{xy}/\sigma^2B,
\label{Rs}
\ee
which measures the ratio of the Hall current and the longitudinal
current squared (see Fig. \ref{R}).  
We note that ${\cal R}_{\sigma} = R_H[1+(\tan\theta_H)^2]$ deviates 
from the ordinary Hall coefficient $R_H$ when the Hall angle $\theta_H$ is large.  
Remarkably, Fig. \ref{R} shows that, below 1 K, ${\cal R}_{\sigma}$ is 
just a constant equal to ${\cal Z}$, even though both $\sigma$ and $\sigma_{xy}$ 
are increasing with strong curvature.  Above 1 K, ${\cal R}_{\sigma}$ deviates significantly
from ${\cal Z}$ as the excess Hall current grows, but only for fields $H<H_Z$.  Above $H_Z$ (arrow), we see that ${\cal R}_{\sigma}$ again settles down to the value ${\cal Z}$.  The constancy of ${\cal R}_{\sigma}$ is direct evidence that both the anomalous 
MR and $\sigma_{xy}$ reflect the enhancement in $\ell$.
As seen in the phase diagram Fig. \ref{kT}a, $H_Z$ (open circles) lies 
significantly below $H_s$.  The simple Hall response determines the high-field Hall behavior 
in the regions I and II.  Hence the complicated Hall response in 
CeCoIn$_5$ arises solely from the excess Hall current 
which is responsible for the weak-field ``knee'', but is suppressed above $H_Z$.
\bfig[t]  
\incl[width=6cm]{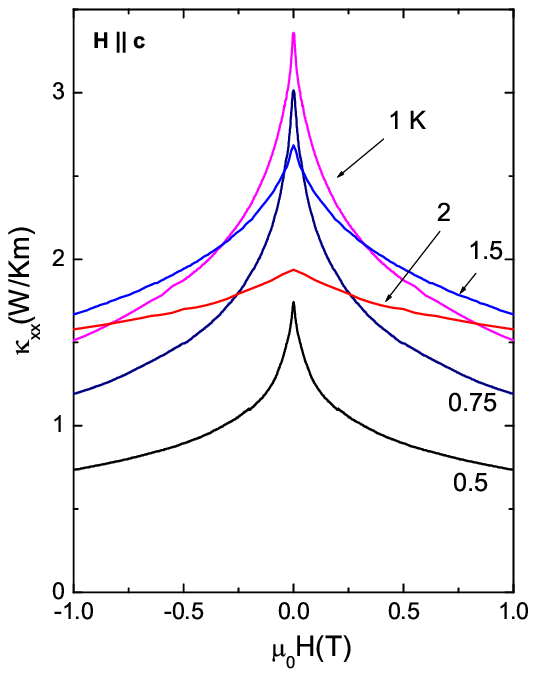}
\caption{\label{klowT}
Expanded view of $\kappa_{xx}\equiv \kappa$ vs. $H$ in the
vortex state of CeCoIn$_5$ ($T<T_c$).
The sharp peaks in weak $H$ arise from the field suppression of the broad peak
in $\kappa(0,T)$ below $T_c$ (see Fig. \ref{kT}b).
}
\efig
\section{Quasiparticles in superconducting state}
These findings have interesting implications for the superconducting state.  
In heavy-electron systems, pairing mediated by the exchange of spin fluctuations 
has been proposed as the likely mechanism for the SC state.  However, the evidence 
to date for spin exchange has been indirect.  Here, we have exploited the 
unusually large strong $H$ dependence of the tensors $\kappa_{ij}$ and $\sigma_{ij}$ to 
show that spin excitations in fact provide the dominant scattering channel at low $T$.  
Insofar as pairing likely arises from the dominant scattering channel, our results 
provide rather direct evidence for spin-mediated pairing.

Finally, we comment on the extraordinary peak in $\kappa$ vs. $H$ that appears
below $T_c$ (Fig. \ref{klowT}).  The sharp reduction of the peak amplitude in $H$
is very similar to the $\kappa$ vs. $H$ curves below $T_c$ in untwinned
$\rm YBa_2Cu_3O_7$ (YBCO)~\cite{Zhang01}.  The extreme sensitivity to $H$
is interpreted as caused by scattering of nodal qp by vortices~\cite{Krishana,Zhang01,Durst}.  
In YBCO, the observation of a large anomaly in $\kappa_{xy}$ that peaks at 
finite field provided early key evidence that
the peak in $\kappa$ arises from enhancement of $\ell$ of nodal excitations
in a $d$-wave superconductor.
Similarly, the low-field peak reported here in $\kappa_{xy}$
at 0.5 and 0.75 K (Fig. \ref{WF}a) confirms that 
the cusp-anomaly in $\kappa_{xx}$ is electronic in origin.
The steep increase in $\ell$ below $T_c$ implies that the nodal qp does not
experience the intense scattering from spin excitations.   
The close similarity between CeCoIn$_5$ and YBCO suggests
that a steep enhancement of $\ell$ below $T_c$ associated 
with nodal quasiparticles may be generic to electronic-mediated 
pairing with $d$-wave symmetry.

\section{Discussion}
We have exploited the unusually large Righi-Leduc effect in CeCoIn$_5$
to determine the Wiedemann-Franz ratio of its charge carriers 
at low $T$.  As shown in Fig. \ref{WF}, the 2 Hall conductivities 
$\kappa_{xy}$ and $\sigma_{xy}$ are strongly non-linear, with a 
change-in-sign of the curvature occuring as $H$ increases from 0 to 12 T.
Remarkably, over a broad interval of $T$, the 2 quantities may be scaled together using an $H$-independent Lorenz parameter ${\cal L}_H(T)$. 
We find that ${\cal L}_H$ is weakly $T$ 
dependent and close to the Sommerfeld value $(\pi^2/3)(k_B/e)^2$.  
The strict insensitivity of ${\cal L}_H$ to field allows the electronic
heat conductivity $\kappa_e$ to be determined unambiguously.  
On subtracting $\kappa_e$ from the observed $\kappa_{xx}$, 
we uncover a large background charge-neutral term $\kappa_b$ that 
is field sensitive.  As $H$ increases, $\kappa_b$ falls 
while $\sigma$ rises in proportion.  This implies that 
spin excitations are the dominant scatterers of the electrons.  

The transport picture that emerges is that, throughout region I in zero $H$
(Fig. \ref{kT}a), the electrons are strongly scattered by spin excitations.  
Moreover, the spin excitations contribute 
the dominant share of the charge neutral thermal conductivity $\kappa_b$,
which accounts for $\sim 45\%$ of the observed $\kappa_{xx}$ at $T_c$.
In a finite $\bf H||c$, the density of spin excitations is strongly
suppressed.  This leads to 2 correlated effects.  The neutral heat term
$\kappa_b$ is suppressed, while the 3 electronic currents $\sigma$, $\sigma_{xy}$
and $\kappa_e$ grow in proportion, as a result of strong enhancement
of $\ell$.  The trend in $\kappa_b$ suggests that the full suppression of
spin scattering is attained when $H\rightarrow H_s$.
An interesting scaling relationship between $\sigma^2$ and $\sigma_{xy}$ is found. (Two recent findings related to this work are Refs.~\cite{Singh,Izawa2}.)

\acknowledgments
We thank Y. Matsuda, K. Behnia, 
and L. Taillefer for useful discussions.
Y. O. acknowledges partial support by the Nishina Memorial 
Foundation.  Research at Princeton University and the Brookhaven 
National Laboratory was supported by NSF (DMR 0213706) 
and by the U.S. Department of Energy (DE-Ac02-98CH10886), respectively.


\begin{thebibliography}{0}

\bibitem{Petrovic} {Petrovic C. \etal} {J. Phys. Condensed Matter} {\bf 13}, {L337} (2001).

\bibitem{Movshovich} {Movshovich R. \etal} {Phys. Rev. Lett.} {\bf 86}, {5152} (2001).

\bibitem{Izawa} {Izawa K. \etal} {Phys. Rev. Lett.} {\bf 87}, {057002} (2001).

\bibitem{Ormeno} {Ormeno R. J. \etal} {Phys. Rev. Lett.} {\bf 88}, {047005} (2002). 

\bibitem{Aoki} {Aoki H. \etal} {J. Phys. Condensed Matter} {\bf 16}, {L13} (2004).

\bibitem{FFLO} {Radovan H. A. \etal} {Nature} {\bf 425}, {51} (2003); 
{Bianchi A. \etal} {Phys. Rev. Lett.} {\bf 91}, {187004} (2003);
{Kakuyanagi K. \etal} {Phys. Rev. Lett.} {\bf 94}, {047602} (2005).

\bibitem{Paglione} {Paglione J. \etal} {Phys. Rev. Lett.} {\bf 91}, {246405} (2003).

\bibitem{Bianchi} {Bianchi A. \etal} {Phys. Rev. Lett.} {\bf 91}, {257001} (2003).

\bibitem{Bel} {Bel R. \etal} {Phys. Rev. Lett.} {\bf 92}, {217002} (2004).

\bibitem{Onose} {Onose Y., Li L., Petrovic C. and Ong N. P. } {Europhys. Lett.}
 {\bf 79}, {17006} (2007).
 
\bibitem{Kalychak} {Kalychak, Y. M. \etal} {Russian Metallurgy}
{\bf 1}, {213} ({2007}).

\bibitem{Krishana} Krishana K., Harris J.M. and Ong N.P., {Phys. Rev. Lett.}
{\bf 75}, {3529} (1995).

\bibitem{Zhang01} {Zhang Y. \etal} {Phys. Rev. Lett.} {\bf 86}, {890} (2001). 


\bibitem{Durst} {Durst A. C., Vishwanath, A. \and Lee P. A.}, {Phys. Rev. Lett.} {\bf 90}, {187002} (2003). 

\bibitem{Nakajima}  {Nakajima Y. \etal} {J. Phys. Soc. Jpn.} {\bf 73}, {5} (2004).


\bibitem{Zhang00} {Zhang Y. \etal} {Phys. Rev. Lett.} {\bf 84}, {2219} (2000). 

\bibitem{Paglione2} {Paglione J. \etal} {Phys. Rev. Lett.} {\bf 97}, {106606} (2006).

\bibitem{CeRhIn} {Paglione J. \etal} {Phys. Rev. Lett.} {\bf 94}, {216602} ({2005}).

\bibitem{Tanatar} {Tanatar M. A., Paglione J., Petrovic C., \and Taillefer L.}
{Science}  {\bf 316}, {1320} ({2007}).
 
\bibitem{Broholm} {Broholm C.} \emph{private communication}

\bibitem{1D} {Sologubenko A.V. \etal} {Phys. Rev. Lett.} {\bf 84}, {2714} (2000);
{Hess C. \etal} {Phys. Rev. Lett.} {\bf 90}, {197002} (2003);
{Jin R. \etal} {Phys. Rev. Lett.} {\bf 91}, {146601} (2003).


\bibitem{Singh} {Singh \etal}, {Phys. Rev. Lett.} {\bf 98}, {057001} ({2007}).

\bibitem{Izawa2} {Izawa \etal}, cond-mat/0704.1970.



\end{thebibliography}
\end{document}